\documentclass[10pt,conference]{IEEEtran}
\usepackage{amsmath,amsfonts,amsthm,amssymb}
\usepackage{algorithm,algorithmicx,algpseudocode}
\usepackage{xspace}
\usepackage{graphicx}
\usepackage{textcomp}
\usepackage{xcolor}
\usepackage{listings}
\usepackage{multirow}
\usepackage{pifont}
\usepackage{fontawesome}
\usepackage{hyperref}
\usepackage{subfig,paralist}
\usepackage{pifont}
\usepackage{enumitem}
\usepackage{url}
\usepackage{booktabs}
\usepackage{tabularx}
\usepackage{multirow,array}
\usepackage{diagbox}
\usepackage{bm}
\usepackage{cite}

\usepackage{titlesec}
\titleformat{\subsubsection}[runin]
       {\vspace{0.1cm}\normalfont\fontshape{\itdefault}\fontseries{\bfdefault}\selectfont}
       {\arabic{subsubsection})}
       {0.5em}
       {}
       [\ \ ]

\usepackage{tikz}

\newtheorem*{theorem*}{Theorem}

\usepackage{tcolorbox}
\definecolor{mycolor}{rgb}{0.122, 0.435, 0.698}
\definecolor{gray1}{gray}{0.3}

\definecolor{codegreen}{rgb}{0,0.6,0}
\definecolor{codegray}{rgb}{0.5,0.5,0.5}
\definecolor{codepurple}{rgb}{0.58,0,0.82}
\definecolor{backcolour}{rgb}{0.95,0.95,0.92}
\lstdefinestyle{mystyle}{
    commentstyle=\color{codegreen},
    keywordstyle=\color{magenta},
    numberstyle=\tiny\color{codegray},
    stringstyle=\color{codepurple},
    basicstyle=\tiny\ttfamily,
    breakatwhitespace=false,
    breaklines=true,
    captionpos=b,
    keepspaces=true,
    numbers=left,
    numbersep=5pt,
    showspaces=false,
    showstringspaces=false,
    showtabs=false,
    tabsize=2,
    columns=fixed
}
\newcommand{\said}[1]{%
\begin{tcolorbox}[colframe=mycolor,boxrule=0.5pt,arc=4pt,
      left=6pt,right=6pt,top=6pt,bottom=6pt,boxsep=0pt,width=\columnwidth]%
      {\small\emph{#1}}
\end{tcolorbox}%
}

\definecolor{darkgreen}{rgb}{0.0, 0.5, 0.0}
\definecolor{darkred}{rgb}{0.82, 0.1, 0.26}

\newcounter{examplecounter}
\newcommand{\example}{\refstepcounter{examplecounter}\emph{Example \theexamplecounter{}}.\xspace}

\AtBeginDocument{%
  }

\begin{document}

\title{Fundamental Challenges in Cybersecurity and\\a Philosophy of Vulnerability-Guided Hardening}

\author{\IEEEauthorblockN{Marcel B\"ohme\\MPI-SP, Germany\vspace{-0.4cm}}}

\maketitle
\begin{abstract}
  Research\,in\,cybersecurity\,may\,seem\,reactive,\,specific, ephemeral, and indeed ineffective. Despite decades of innovation in defense, even the most critical software systems turn out to be vulnerable to attacks. Time and again. Offense and defense forever on repeat. Even provable security, meant to provide an indubitable guarantee of security, does not stop attackers from finding security flaws. As we reflect on our achievements, we are left wondering: Can security be solved once and for all?

  In this paper, we take a philosophical perspective and develop the first theory of cybersecurity that explains what \emph{fundamentally} prevents us from making reliable statements about the security of a software system. We substantiate each argument by demonstrating how the corresponding challenge is routinely exploited to attack a system despite credible assurances about the absence of security flaws. To make meaningful progress in the presence of these challenges, we introduce a philosophy of cybersecurity.

\end{abstract}

\section{Introduction}
In the Pwn2Own hacking contest this year, a single person (Manfred Paul) demonstrated successful exploits for all major browsers: Google Chrome, Mozilla Firefox, Apple Safari, and Microsoft Edge \cite{paul}.
Just Chrome is used by 3.5 \emph{billion}~people.\footnote{\scriptsize\url{https://backlinko.com/chrome-users}} Today, thanks to Paul, these major security flaws are fixed. Last year, evidently, most people accessed the internet with the door open to potential attackers---despite decades of research in defensive security, despite an abundance of mitigations,
despite a dedicated red team (Google P0) that is considered the best in the world.
From the outside, 
research in cybersecurity might seem technology-specific, reactive, and ostensibly ineffective. 

But why? Software is entirely virtual and can be described completely \cite{hoare2}: A program's source code is meant to formally express the programmer's intention using the syntactic and semantic rules of the programming language. As the behavior of a software system arises from well-defined instructions, we must be able to formally reason about all its properties. Surely there exists an approach that will forever guarantee the security of our systems. Only, we have not found it, yet?
\vspace{-0.1cm}
\said{Fundamentally, can we guarantee the absence of security flaws? If not, what precisely prevents us from making reliable statements about the security of a software system?}
\vspace{-0.1cm}

This question about the effectiveness of program analysis is at the very heart of \emph{Software Engineering}. Yet, our battle-tested research methods are not well equipped to answer it. Since we do not propose a new approach or intervention, we can hardly run measurements or experiments to gain insight. Since we look for guidance on the science of security itself, we cannot expect answers from descriptive methods, either.

In this paper, we take an objective, philosophical perspective on the effectiveness of cybersecurity tooling and develop a \emph{theory of cybersecurity} that explains why we must, in general, assume that exploitable security flaws exist in any sufficiently large software system. We substantiate each argument by demonstrating how the corresponding challenge is routinely exploited to attack a system despite credible assurances about the absence of security flaws.
We believe, now is the time to reflect on our scientific progress more fundamentally and to shed light on the limitations of our approaches to reason about the properties of a system. We do not claim, our theory is final or complete. Instead, we expect it to be discussed, improved, and refined in a protracted scientific~debate.

How do we make meaningful progress? The quest for truth in the absence of a mechanism guaranteed to arrive at the truth has existed since the time of enlightment. In his philosophy of science, Popper suggests, scientific theories evolve in a counterexample-guided manner (\emph{conjectures and refutations})~\cite{popper}. In his philosophy of maths, Lakatos suggests, even theorems evolve in a counterexample-guided way (\emph{proofs and refutations})~\cite{lakatos}. Similarly, in our philosophy of security, we suggest, security claims evolve via counterexamples, too.

Consider fishing as a metaphor for bug finding. A \emph{fishernet} represents our tools and processes while the \emph{fishes} represent the security flaws in our software systems. Our theory explains why, for every  net, there will always be a fish that slips through. Clearly, this does not undermine the utility of the net. Our philosophy turns the objective from developing the ultimate fishernet to an incremental, counterexample-guided evolution of our nets maximize effectiveness empirically.

Concretely, our philosophy suggests a \emph{vulnerability-guided hardening} of our security approaches. We cannot assume that an approach $T$ developed to protect software system $S$ is final in any way. Instead, we recommend
(i)~to find some evidence $c$ \emph{against} the security of $S$ despite $T$ (e.g., by exploiting the challenges we identified),
(ii)~to patch $S$ and $T$ so as to account for $c$, and
(iii)~to restart by finding the next evidence $c'$. Later, we discuss a concrete application to software verification.

Like counterexample-guided approaches in science and maths, vulnerability-guided hardening addresses the identified fundamental limits \emph{empirically}. In contrast to science and mathematics, we can support this method with automation. For instance, we should invent novel debugging techniques for our approaches $T$ that allow us to pinpoint the precise root cause of their failure to discover or mitigate a given vulnerability $c$.
We need innovative techniques to programmatically reconcile the discovery of a vulnerability $c$ to improve the approach $T$.


\vspace{0.1cm}
\noindent
In summary, this paper makes the following contributions:
\begin{itemize}[leftmargin=0.3cm,itemsep=0.1cm]
  \item We introduce a \emph{theory of cybersecurity} which explains why we must assume that exploitable security flaws exist in any sufficiently large software system. We discuss nine (9) fundamental challenges for software security analysis and demonstrate how they are routinely exploited.
  \item We introduce a \emph{philosophy of cybersecurity} which serves as guiding principle granting the fundamental insecurity of our software systems.\footnote{To clarify, a philosophy is really just a specific way of thinking about a problem. It represents a particular perspective on the nature of that problem.} We suggest that a security mechanism cannot be deemed effective \emph{generally} and \emph{with finality}, e.g., if it performs well on a benchmark. Instead, we must incrementally find evidence \emph{against} the generality of our security mechanisms by subjecting them to a method of systematic scrutiny, we call vulnerability-guided hardening.
  \item We introduce \emph{meta verification} to instantiate our method of vulnerability-guided hardening for software verification and to tackle the long-standing crux that formal guarantees for empirical systems are empirically unreliable \cite{herley,demillo,secdev18}.
\end{itemize}

\section{Background}
\subsection{Science of Security}
In pursuit of more rigorous foundations, there have been many calls for making the research discipline of cybersecurity more scientific~\cite{geer,blakely,evans,peisert,xumoti}. For instance, Schneider \cite{schneider} argues 
that the cybersecurity research community ought to construct a body of laws for predicting the consequences of design and implementation choices. The laws should ``(i)~transcend specific technologies and attacks, yet still be applicable in real settings, (ii)~introduce new models and abstractions, thereby bringing pedagogical value besides predictive power, and (iii)~facilitate discovery of new defenses and describe non-obvious connections between attacks, defenses, and policies, thus providing a better understanding of the landscape''.

Herley and Van~Oorschot~\cite{herley} provide an excellent survey of such calls for a more scientific approach and ponder what it means for security to be ``more scientific''. They start with~a~comprehensive survey of view points in Philosophy of Science and
discuss concrete opportunities for cybersecurity generally.\footnote{Herley and Van~Oorschot~\cite{herley} also critize how empirical work in security aims to \emph{verify} existing beliefs while it really should falsify them to disambiguiate possibilities and suggest new theories. Furthermore, referring to defensive advice like ``effective passwords must have a certain entropy", they criticize the advancement of unfalsifiable claims and the conflation of unsupported assertions and argument-by-authority with evidence-supported statements.} On our topic of mechanisms for reliable statements about the security of a software system, they discuss \emph{provable security} specifically, i.e., techniques that use the tools of logic, formal methods, cryptography, and mathematics to derive a formal guarantee about the security of a system.\footnote{Including cryptography, model checking, software verification, verified compilation, proof-carrying code, and (exhaustive) symbolic execution.}

Provable security is \emph{not technically} a science in the same way as mathematics is not technically a science, the authors argue~\cite{herley}. While a \emph{science} seeks to make inductive statements from observations about the empirical world (which requires hypotheses to be falsifiable), \emph{provable security} seeks to make deductive statements from axioms within a formal system. This observation is further explored by Murray and Van~Oorschot \cite{secdev18} who expose challenges of interpreting the formal guarantee for the actual system (which is labeled as ``formally verified'') and of enforcing the formal assumptions.

Herley and friend~\cite{herley} call this the \emph{inductive-deductive split}: ``Speaking of mathematical guarantees as if they are properties of real-world systems is a common error. [..] It is worth being unequivocal on this point. There is no possibility whatsoever of proving rigorously that a real-world system is [..] invulnerable to (all) attacks'' \cite{herley}.
In this paper, we make this assertion more precise by identifying concrete challenges (beyond provable security), but also fundamentally resolve this ``inductive-deductive split'' which Herley and Van~Oorschot consider as insurmountable. We agree that formal claims are not guaranteed to hold for the empirical system but suggest that the distance between the formal and the empirical world is iteratively reduced via (empirical) counterexamples by applying the scientific method to the \emph{process} of provable security itself (inspired by Lakatos' philosophy of mathematics \cite{lakatos}).

While there exist many approaches to analyse the security of a software system, only provable security provides a \emph{formal guarantee}. The concrete assumptions, the concrete properties, and the formal reasoning framework are {explicitly}, {precisely}, and {formally} spelt out. A proof gives a formal and universal guarantee that \emph{these} properties hold within \emph{this} model of the software system if \emph{these} assumptions are met.

Koblitz and Menezes~\cite{anotherlook,anotherlook3,provsec2} critizise the finality with which these guarantees are advertised. They provide examples of incorrect proofs, non-constructible algorithms, idealized models, broken assumptions, and inadequate definitions in  cryptography.
Indeed, Lakatos \cite{lakatos,sep-lakatos} argues that proofs of mathematical theorems aren't final, either. Rather, he suggests, these water-tight deductions from well-defined premises are the perhaps temporary {end-points} of an evolutionary, {dialectical} process. De~Millo et al. \cite{demillo} observe that mathematics evolves within a \emph{social process} where the mathematicians' confidence in the validity of a theorem increases as it becomes subject to more scrutiny; they assert that software verification is ``bound to fail'' precisely because of the lack of this social process.
In this paper, we build on Lakatos' philosophy of mathematics to develop a philosophy of security. We explain how De~Millo's social process may not be required, after all, and suggest how it may be substituted by a general, systematic counterexample-guided method.

\said{``The aim of program verification [..] is to increase dramatically one's confidence in the correct functioning of a piece of software. [..] Contrary to what its name suggests, a proof is only one step in the direction of confidence. We believe that, in the end, it is a social process that determines whether mathematicians feel confident about a theorem--and we believe that, because no comparable social process can take place among program verifiers, program verification is bound to fail.''\vspace{-0.1cm}
\begin{flushright}---\emph{De Millo, Lipton, and Perlis \cite{demillo}}\end{flushright}}

\subsection{Generating Reliable Statements about Software Security}
\subsubsection{What is a secure software system?}
An attacker should not be able to read what they shouldn't (\emph{confidentiality}), to {write} what they shouldn't (\emph{integrity}), or to disturb the service of others (\emph{availability}).

\subsubsection{Attacks.}
Today, the largest class of security flaws is due to \emph{undefined behavior} and \emph{memory unsafety} more specifically. This includes buffer overflows, heap/stack smashing, return-oriented programming (ROP), return2libc, and format-string vulnerabilities. Other attacks exploit \emph{data races}, including time-of-check time-of-use (TOCTOU) vulnerabilities, or \emph{hardware defects} such as RowHammer \cite{rowhammer}.

Several types of attacks are related to problematic information flow. An \emph{injection attack} is a problematic information flow from a public source to a sensitive sink that might threaten integrity. This includes command injection, code injection, SQL injection, Log4Shell (JNDI) \cite{log4j2}, Cross Site Scripting (XSS), and deserialization vulnerabilities. An \emph{information leakage} is a problematic information flow from a sensitive source to a public sink that might threaten confidentiality. This includes hardware and software side channel vulnerabilities \cite{spectre}.

Many attacks are domain-specific. There are hypocrite commit vulnerabilities and compiler backdoors for in \emph{supply chain security}. There are man-in-the-middle (MitM) and spoofing attacks in \emph{network security}. There are click jacking and XSS attacks in \emph{browser security}. There are jailbreaks and communication-stack-based attacks in \emph{mobile security}. There is privilege escalation in \emph{OS and hypervisor security}.

\subsubsection{Defenses.}
The best defense is \emph{education and training}. To make our systems more secure, we should teach developers adversarial thinking \cite{adversarialThinking} as well as offensive and defensive strategies in hands-on labs and capture-the-flag (CTF)-type competitions. At the university-level, students should be able to take a long-term, technology-unspecific, automation-centric perspective on software security in-the-large.

Defenses (or mitigations) require us to draw a line between what an attacker \emph{must never} be allowed to do (read, write, or disturb) and what the user \emph{must always} be allowed to do. For instance, the \emph{principle of execution integrity} suggests to establish programmer's intention and to avoid deviations. For memory safety, this includes control-flow integrity (CFI) where the intended control-flow is statically precomputed and dynamically enforced, but also canaries, shadow stacks, and fat pointers. Language-based security requires intention to be made explicit, e.g., via borrow checking.

The \emph{principles of isolation} and \emph{least privilege} require us to specify isolation primitives and authorization policies. For memory unsafety, memory is segmented and guard pages introduce, components are compartmentalized, and pointers signed.
For data races, concurrent processes are isolated using locks and semaphors.
For injection attacks, sensitive sources are isolated from public sinks.
For information leaks, public sources are isolated from sensitive sinks.
For attacks on operating systems, processes are assigned to rings or sandboxes, and privileges / permissions are assigned to users.

\begin{figure}\centering
  \includegraphics[width=0.6\columnwidth]{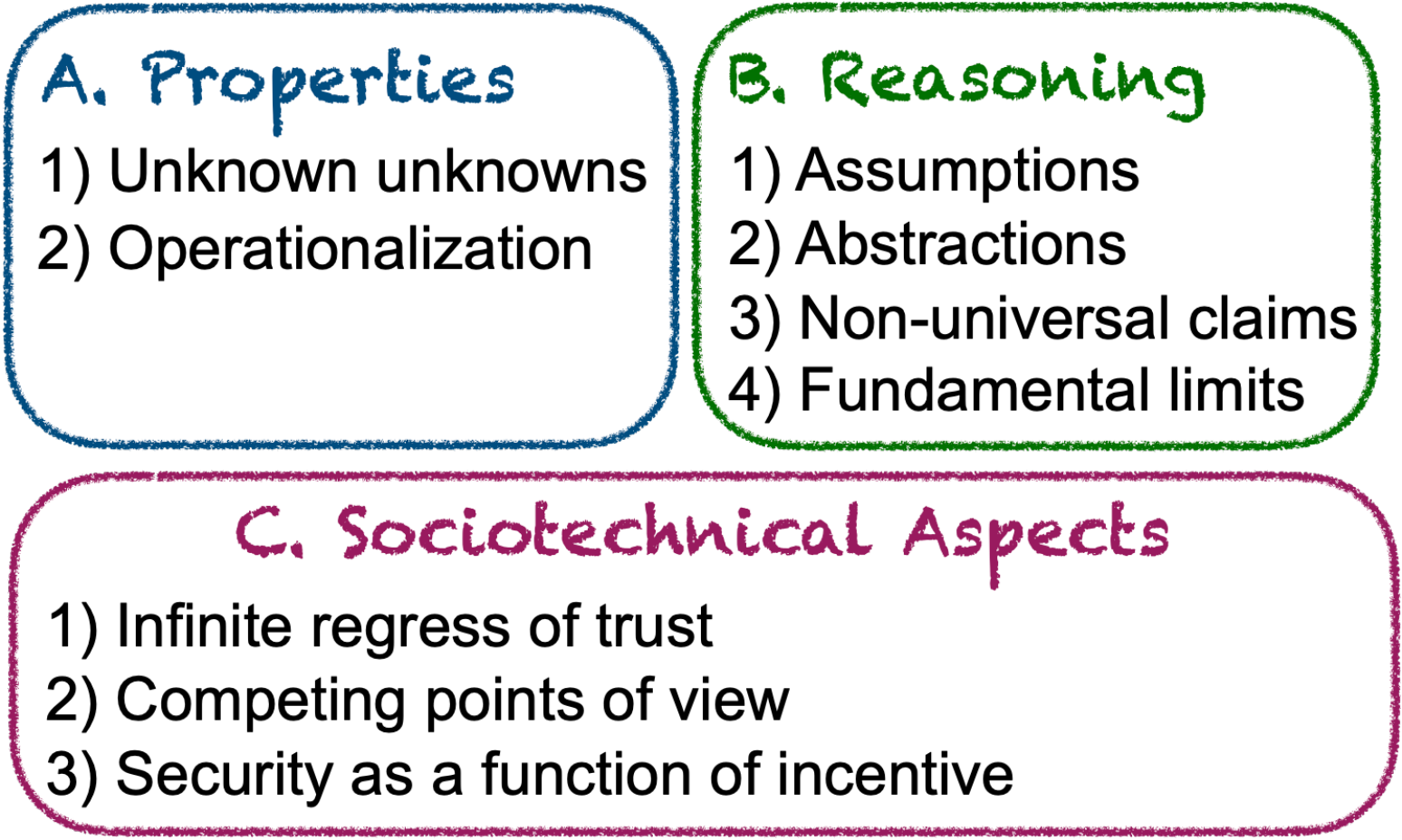}
  \caption{Overview of the identified challenges to make reliable statements about the security of a software system.\vspace{-0.2cm}}
  \label{fig:overview}
\end{figure}

\section{A Theory of Guarantees in Cybersecurity}
\label{sec:challenges}
We develop a first theory that explains why we must, in general, assume that security flaws exist in any sufficiently large software system. Our theory elucidates specific blindspots in our approaches to reason about security and how they can be exploited by an attacker. As evidence for our theory, for each blindspot, we discuss concrete examples where that blindspot was exploited, often despite credible assurances about the absence of security flaws.
We distinguish these challenges in three categories (cf. Fig.
\ref{fig:overview}): (A)~the \emph{security properties} we need to establish and operationally define, (B)~the \emph{reasoning system} (or security mechanism) that we use to check or enforce these properties, and (C)~the \emph{sociotechnical challenges}.

\subsection{Challenges to Distinguish Secure from Insecure Software}\label{sec:oper}
We cannot perfectly recognize and precisely describe the distinction between secure and insecure software.

\vspace{0.1cm}
\noindent
On the one hand, we agree that an attacker should \emph{not} be able
\begin{itemize}
  \item to read what they should not read (confidentiality),
  \item to write what they should not write (integrity), or
  \item to disturb the service for others (availability).
\end{itemize}
For instance, a cryptographic system whose execution time depends on the secret key violates the confidentiality criterion. For the cryptographic system to be secure, the constant-time property must hold (amongst other properties).

\textbf{\emph{Drawing the line}}. On the other hand, we may not recognize \emph{all} properties that need to hold for the system to be secure (unknown unknowns). Even if we did, we may \emph{not} be able to describe those properties precisely enough to detect, prevent, or mitigate \emph{all instances} of their violation (operationalization).

\subsubsection{Which line? Unknown unknowns.}\label{sec:swan}
The first fundamental challenge is to know \emph{which} properties should need to hold for a system to be secure. We may be absolutely convinced that no security flaws exist---until some behavior, that is problematic only in retrospect, is exploited by an attacker. We call such ``unknown unknowns'' as \emph{black swan properties}.

\example
For instance, Nakamoto's consensus protocol has been formally shown to \emph{``guarantee safety and liveness"} \cite{nakamoto}. However, there is nothing preventing malicious participants from violating their part of that protocol and casting their vote more than once. To account for this, we need a secure consensus protocol to have another property called \emph{accountable safety} \cite{dilemma}. Moreover, there is nothing preventing an attacker from creating network partitions such that the block confirmation procedure cannot be relied upon. To guarantee safety under network partitions, we need a secure consensus protocol to have another property called \emph{finality} \cite{ebbflow}. Many other properties need to be satisfied for a consensus protocol to be able to withstand attacks, some of which are known while others remain unknown until their absence is exploited.

\example
In 2018, Google introduced Android 9 Pie with a tool called Markup that allowed users to crop images.
In 2023, Security researchers found that this crop is only virtual  (CVE-2023-21036). The cropped part of the image is only \emph{marked} as cropped. The data remains. This violates confideniality because the user might have shared an image where sensitive information was cropped, e.g., on an image of a credit card. Today, we know that Markup should ensure that cropping \emph{actually} removes the cropped image data.

\example{\label{ex:sel4}}
Eleven years ago, Murray et al. \cite{sel4} gave the first proof of information flow security for an industrial-strength, high-performance operating system microkernel, called seL4. The formal demonstration of confidentiality and integrity with respect to the given security policy is considered a milestone in software verification. SeL4 was written in 8.7k lines of code and verified with 200k lines of proof. For this milestone, the team received the well-deserved 2022 ACM Software System award.
Four years later, Spectre (CVE-2017-5753) was shown to violate the confidentiality guarantee. The speculative execution of secret-dependent branches yields small differences in execution timing which allows an attacker to infer the secret from observations of the execution time.
Since then we know that seL4 must also satisfy the constant-time property to guarantee confidentiality.

\emph{Mitigation} (threat modeling). Some program behaviors are security flaws only \emph{in hindsight}. What exactly makes a system vulnerable to attacks may be difficult to anticipate beforehand.
We can attempt to anticipate certain kinds of properties by following a systematic elicitation process. For instance, we can define a \emph{threat model}, i.e., a structured representation of all the information that affects the security of an application. Some attacks will be considered within the threat model while others will be considered outside.
However, some attacks will not be considered by the threat model in the first place (as our examples demonstrate).

\subsubsection{Drawing the line (operationalization).}\label{sec:line}
Even if we know which high-level property should hold, we must define that property concretely and operationally to be able to mechanically check or enforce it. An attacker can violate a high-level property while keeping its operationalization in tact.

\example
We know that seL4 must satisfy the constant-time property to guarantee confidentiality from Example~\ref{ex:sel4}. How do we formalize this property? We could enforce that there are no secret-dependent branches that are speculatively executed to mitigate Spectre (CVE-2017-5753). We could enforce that the number of cache hits and misses is not secret-dependent to mitigate Meltdown (CVE-2017-5754). However, this is only a subset possible secret-dependent optimizations that need to be handled, today and in the future \cite{lmtest}. To recover the formal guarantee of confidentiality, the seL4 project would need to formalize \emph{all} of these operational definitions.

\example
To prevent an attacker from executing arbitrary code, we could blocklist specific system calls that are known to directly or indirectly facilitate arbitrary code execution.\footnote{For instance, FlawFinder (\url{https://dwheeler.com/flawfinder}) is used to audit calls to strcpy, strcat, sprintf, chown, mktemp, exec*, system, and popen.} However, this could also block important system calls that are required for many programs to function properly, and we might forget to block other system calls that could still be exploited. In fact, there are many other ways to gain arbitrary code execution instead of system calls. We must decide concretely how we operationally ``encode'' that attackers should not be able to execute arbitrary code. Bypassing our operationalization, an attacker may still execute arbitrary code.

\emph{Drawing the right line}. Many types of security flaws require an explicit decision between what an attacker \emph{must never} be able to do versus what a user \emph{must always} be able do. For instance, \emph{injection vulnerabilities}, such as command/SQL/XSS injection (Log4Shell \cite{log4j2}) or deserialization vulnerabilities require input sanitization (e.g., via regular expressions), block- or allow-lists, or a list of sensitive sinks. Similarly, \emph{information leakage vulnerabilities} (e.g., a credit card number leaking into a shared log) require lists of sensitive sources and public sinks.
Because we always need to settle on a \emph{specific} operational definition, an attacker may be able to violate the high-level property while keeping the operationalization in tact.

\emph{Using the right pen}.
The process of operationalization is limited by the \emph{language} (or theory) available to encode that property. For instance,
  \emph{pattern-based static analysis}\footnote{A \emph{static analysis} finds bugs \emph{without} executing the program by analyzing the source code or compiled binary.} cannot find security flaws that cannot be expressed by the pattern language (e.g., as extended regular expression).\footnote{Weggli\,\url{https://github.com/weggli-rs}, SemGrep\,\url{https://github.com/semgrep}, {FlawFinder} \url{https://dwheeler.com/flawfinder/}.}
  \emph{Semantics-based static analysis} cannot find security flaws inexpressible as assertions on the modelled program state.
  For instance, symbolic execution\footnote{Klee \url{https://klee-se.org/}, Java Pathfinder \url{https://github.com/javapathfinder}.} requires assertions expressed in the given Satisfiability Modulo Theory (SMT).
  Logics-based analysis\footnote{Infer \url{https://fbinfer.com/}, CodeQL \url{https://codeql.github.com/}.} requires assertions expressed in the given logic (e.g., separation logic).
  Software verification\footnote{CompCert \cite{compCert}, seL4 \cite{sel4}, Project Everest \cite{evercrypt,hacl}} requires assertions expressed in the given theory (e.g., Linear Temporal Logic~(LTL)).
  Constructs like pointer or floating point arithmetic, stack/heap, or string/array operations need to be modelled explicitly.
  In fact, papers introducing side channel attacks often come with their own logic to encode such type of attacks \cite{sunjay,guarnieri}.


\emph{Problem of Translation} \cite{lakatos}. How do we know the terms \emph{inside} the operational definition have the same meaning as the terms \emph{outside}? In philosophy, this is the Problem of Translation. For instance, how do we know that \emph{your} definitions of ``bias'' and ``fairness'' are equivalent to opertional definitions that researchers have developed in machine learning \cite{fairness}?

\example
Greenman et al. \cite{ltlmisconceptions} studied how researchers and learners understand LTL properties and found that the encoding of theorems in English language as LTL formulas "was fraught with errors and provides evidence for a large number of misconceptions".
Murray and Oorschot \cite{secdev18} compare the user's interpretation of the formal statement to reading the fineprint in legal contracts.
For instance, they report that even a leading researcher in the field, Benjamin Pierce, required a full week to understand the functional correctness proof of SeL4 deeply enough to prepare two lectures about them in a graduate seminar \cite{secdev18}.

\emph{In summary}, it is difficult to draw a concrete line between a secure and an insecure system precisely enough to prevent an attacker from exploiting security flaws despite concrete and credible assurances about the absence of security flaws.


\subsection{Challenges Of Reasoning About Cybersecurity}\label{sec:reason}
Assuming we could draw the line between secure and insecure systems precisely enough, we are still fundamentally limited in the way we reason about the behaviors of a given software system with respect to that line.
To reason about the behaviors of a software system, we need to \emph{model} its behaviors. From properties of the model, we make claims about properties of the running system. An attacker can attack the system while keeping properties in the model in tact.

The most powerful security approaches reason over some model of the behavior of the software system\begin{itemize}[leftmargin=0.3cm,itemsep=0.1cm]
  \item \emph{Provable security} uses the tools of logic,
  formal methods, and mathematics to derive a
  formal guarantee about the security of a software system. This includes cryptography, model checking, protocol analysis (e.g., consensus or internet protocols), secure-by-construction \cite{synthesis}, proof-carrying code \cite{necula,vanague}, and software verification \cite{piessens}.
  \item \emph{Static analysis} (semantics-based), like symbolic execution,  abstract interpretation, and logic-based static analysis (e.g., Infer, CodeQL, SonarQube, and FindBugs) translates the program's source code (or binary) into a model of computation represented in the available semantics, logic, or theory to check assertions about the behaviors of the system \cite{lipp}.
  \item \emph{Dynamic analysis}, including fuzzing, compartmentalization, sandboxing, trusted execution environments, model the current execution at some (fixed) level of abstraction \cite{sandbox,sandboxescape,sgx,compartment1,compartment2,compartment3, exploitbrokers}.
  \item \emph{Secure-by-design} refers to techniques to avoid introducing security flaws at the design stage. This includes software engineering efforts, such as ensuring security best practices or \emph{threat modelling} to elicit project-specific security requirements as well as \emph{language-based security} efforts to entirely avoid large classes of security faults, such as memory safety issues, at the OS or programming language level \cite{secureSE,stride}.

\end{itemize}

As we always reason about a \emph{specific} model of the system, an attacker may be able to violate a security property of the real system while keeping properties in the model in tact. For instance, the attacker may violate an assumption of the model, exploit a security flaw at a lower level of abstraction, or exploit flaws that are missed in the model.

\subsubsection{Assumptions.}\label{sec:assumptions}
When reasoning about a software system, we make assumptions about the actual, system as it is running in production. An attacker can violate an assumption. These assumptions are often about matters in the empirical world \emph{outside} the model used to reason about the deployed system.

\example
The security of cryptographic protocols depends on \emph{assumptions about the computational resources} of an attacker or the computational complexity of a mathematical problem. For example, the RSA public key encryption is secure assuming that we cannot efficiently factor large numbers into their prime factors. This assumption is broken on a quantum computer with Shor's algorithm. Cryptography, as maturing field with important innovations, has a long history of make-and-break cycles where assumptions are exploited to break protocols otherwise provably secure \cite{provsec,denning99,herley}.

\example
Assumptions (or axioms) are fundamental in provable security.
Bognar et al. \cite{piessens} analyzed two embedded trusted-execution architectures that were formally shown \emph{provably-secure} \cite{sancus,vrased}. They identified nine assumptions, e.g.,
  (i)~Enclave software cannot access unprotected memory or manipulate interrupt functionality.
  (ii)~Interrupted enclaves can only be resumed once with \texttt{reti} and not be restarted from ISR.
  (iii)~The \texttt{dma\_addr} bus contains the full address.
  (iv)~All components use a consistent key size.
For every assumption, Bognar et al. demonstrated how it could be exploited to {successfully attack the trusted architectures}.
%
Kang et al. \cite{sepcomcert} analyzed a compiler that was formally shown \emph{provably-correct} \cite{compCert}. They identified an axiom that was empirically invalid and could be exploited such that the compiled binary would compute an unexpected result.

\example
In formal reasoning, we can make assumptions about the ``model-external'' world explicit via \emph{formal contracts}. For instance, \emph{proof-carrying code} \cite{necula} makes explicit which properties otherwise untrusted code satisfies for the execution to be secure. Vanegue \cite{vanague} showed, using weird machines, how these guarantees can be bypassed. \emph{Execution} and \emph{leakage contracts} \cite{leakagecontract} make assumptions about the hardware explicit (to enforce an operational definition of constant-time).

\example
In static analysis, we might make assumptions about the callers of a function if our analysis is \emph{intra}-procedural. We might make assumptions about the maximum number of loop iterations or the maximum depth of recursion. When not all code is available, such as for third-party libraries or system calls, we might choose to model those. We might make assumptions about the maximum size of the inputs or the operational distribution \cite{pse} for a software system.

\example
Assumptions about components outside the model are also made in \emph{language-based security}. For instance, we assume that calls to \texttt{unsafe} in memory-safe languages like Rust or Java, e.g., to interact with the hardware or issue calls to a native C-library, are safe-by-developer. We might also make inconsistent assumptions \emph{across} the language boundary. For instance, Mergendahl et al. \cite{crosslang} demonstrated an attack where control-flow integrity (CFI) is enforced in the C part and memory-safety in the Rust part by corrupting memory in C and hijacking control flow in Rust.

Even if assumptions are stated explicitly, it is sometimes difficult to assess (a) the degree to which an assumption holds for the deployed software system and (b)~the resulting threat to the high-level security guarantee.

\subsubsection{Abstractions.}\label{sec:abstractions}
A central idea in computer science, \emph{abstraction} allows us to focus attention on details of greater importance by removing or generalizing physical, spatial, or temporal details or attributes of real-world objects or systems. However, because we must always reason at an (arbitrary but) \emph{specific} level of abstraction, an attacker may be able to violate a security property at a \emph{lower level} of abstraction even if the property was shown to hold at the higher level of our model.

\vspace{0.1cm}
$\bullet$ \emph{Specification vs implementation}.
There is a gap between a system's {specification} and its {implementation}. Despite demonstrating a security property for the {specification}, an attacker might still be able to violate it in the {implementation}.

\example
SHA-3 is currently considered as the most secure cryptographic hash function. Yet, a simple buffer overflow in the most widely used implementation of SHA-3 allows an attacker to bypass the security by computing (second) preimages and even to execute arbitrary code on the victim's machine \cite{sha3oob}.

\vspace{0.1cm}
$\bullet$ \emph{Source code vs compiled binary}.
There is a gap between the developer-provided source code and the compiler-provided program binary for that code. Given only the source code, without assumptions about the compiler, it is hard to make reliable statements about properties of the executable.

\example
Effectively all memory safety and undefined behavior issues emerge from this abstraction gap.
In the presence of behavior where the language standard imposes no requirement (i.e., undefined behavior), the compiler is allowed to do anything it chooses, including ``having demons fly out of your nose".\footnote{\scriptsize\url{https://groups.google.com/g/comp.std.c/c/ycpVKxTZkgw/m/S2hHdTbv4d8J}} In fact, undefined behavior includes memory safety issues, which currently constitute 80\% of security vulnerabilities exploited in the wild.\footnote{\url{https://googleprojectzero.blogspot.com/p/0day.html}} Undefined behavior also includes type confusion vulnerabilities where the same unchanged variable is interpreted to have different values in different parts of the program.\footnote{\url{https://blog.trailofbits.com/2022/11/10/divergent-representations/}}
%
If the developer decides to handle an instance of undefined behavior, e.g., by checking for an integer overflow in a saturated increment, the compiler might remove the check (because it is itself undefined behavior) but keep the original undefined behavior unhandled.\footnote{\url{https://research.swtch.com/ub}}

\example
Even if the compiler guarantees \emph{semantic equivalence} between source code and executable, it might be possible to \emph{bypass}, in the executable, {security properties} established at the source level. D'Silva et al. \cite{corSec} identified several compiler optimizations, such as dead store elimination, code motion, and function call inlining that are formally guaranteed to be semantic preserving but might (i)~introduce information leaks through persistent state, (ii)~eliminate security-relevant code, or (iii)~introduce side channels. D'Silva et al. explain that "{the operational semantics used in correctness proofs includes the state of the program, but not the state of the underlying machine}" \cite{corSec}. When reasoning about a program's behavior, the tool's abstract model may not reflect these detailed compiler choices.\footnote{\url{https://news.ycombinator.com/item?id=39191507}}

\vspace{0.1cm}
$\bullet$ \emph{Program vs process}.
Finally, there is a gap between the {program} as it is stored in memory and the process as it runs on the machine. Given only the program, without assumptions about the machine, it is hard to make reliable statements about properties of the running process.

\example
Effectively all side channels emerges from this gap.
Even if we can perfectly guarantee the absence of timing side channels by carefully analyzing the program (binary), an attacker might exploit processor-specific optimizations or microarchitectural features like speculative execution to violate the constant time property \cite{spectre,guarnieri,lmtest}.

\example
All hardware-specific software vulnerabilities and hardware-assisted software security emerges from this gap, as well. For instance, even if we can guarantee information flow security in the program \cite{sel4}, an attacker might exploit defective DRAM memory modules (RowHammer) \cite{rowhammer,secdev18} or a bug in the CPU's microcode \cite{amdzen2} to leak sensitive information. The Zenbleed bug allowed anyone who could execute a certain sequence of instructions (e.g., in a sandbox, container, or separate process) on an AMD Zen 2 class processor to read parameters or return values of sensitive functions like strlen, memcpy, and strcmp \emph{anywhere} on that physical machine.

\example
It is convenient for the assembly programmer or the compiler engineer to think of the CPU as a blackbox that implements a specific instruction set, like x86. To study this abstraction, Domas \cite{hardwareTrust,hardwareTrust2} implemented a tool to exhaustively search the x86 instruction set that is actually \emph{implemented} and found critical x86 hardware glitches, previously unknown machine instructions, ubiquitous software bugs, and flaws in enterprise hypervisors.

\emph{In summary}, if we reason about an abstraction of the software system to make claims about the actual, deployed system, an attacker might violate the security property at a lower level of abstraction while keeping the guarantees at the higher level in tact \cite{bratus,dullien,vanague}.
Or as Balakrishnan and Reps put it: ``What you see is not what you execute" \cite{wysinwyx}. We also note that the specific level of abstraction is often fixed by the language (or logic or theory) that is used in the reasoning framework.

\vspace{0.1cm}
\subsubsection{Non-Universal claims.}\label{sec:nonuni}
An under-approximate analysis only allows existential statements.
Approaches like fuzzing, software testing, symbolic execution, runtime verification, or incorrectness logic only allow us to reason about an (observed) subset of all executions. Even if we assume that we know which security properties should hold and how to operationalize them, these approaches do not allow us to make universal claims about the operationalized properties.

\example
Despite years of fuzzing the Suricata open-source intrusion detection and prevention system, a one-line division-by-zero bug that could cause denial-of-service was found only by a careful manual audit.\footnote{ \url{https://twitter.com/catenacyber/status/1646860408014118913}}

\subsubsection{Fundamental limits.}\label{sec:limits}
There are questions about the operational security properties of a software system that we can simply not answer within any sufficiently powerful reasoning framework. This includes approaches in provable security and in semantics-based static analysis.
\begin{itemize}[leftmargin=0.3cm,itemsep=0.1cm]
\item From a math perspective, according to G\"odel's incompleteness theorem, for any sufficiently complex deductive system that is consistent, there are always true statements that cannot be proved in that system. In other words, a reasoning framework cannot be both, consistent and complete.
\item From a verification perspective, according to Rice's theorem all non-trivial semantic properties of programs are \emph{undecidable}. A property is non-trivial if it is neither true for all programs, nor false for all programs.
\item From a program analysis perspective, according to Landi \cite{landi,undecidable} there exists no algorithm that---in general and even under simplifying assumptions---could decide whether there exists an execution of the program where two pointers point to the same memory address.
\item From a security perspective, according to Cohen, there exists no algorithm that, in general, could detect malware \cite{cohen}.
\end{itemize}

\vspace{0.1cm}
\noindent
In practice, theoretical limits have never stopped us from making progress.
We allow our analysis to be over- or under-approximate.
An over-approximate analysis (e.g., abstract interpretation) reports security flaws that do not exist while an under-approximate analysis (e.g., symbolic execution) misses security flaws that do exist.
However, security is a universal claim, and we cannot be sure about the degree to which our heuristic choices impact the analysis result. Worse, we currently do not even know how to \emph{quantify} the loss of guarantee \cite{statistical}. Even if the analysis is sound, an over-approximate analysis may yield so many false positives that the developer might miss the actual security flaw like a needle in a haystack.

\subsection{Sociotechnical Challenges}\label{sec:socio}
\subsubsection{Infinite regress of trust.}\label{sec:trust}
We seek reliable statements about a software system because we do not \emph{trust} its security in the first place. Yet, we trust whatever the security of the system depends on (i.e., the software supply chain, the code review process, the build infrastructure, or the hardware, hypervisor, and operating system that runs our system). Even if we do not, the same argument applies recursively. An attacker may attack the software system by exploiting bugs or security flaws in whatever we trust to work as expected.

\example
Cryptography ensures the confidentiality of some data (e.g., messages) \emph{only if we trust} the confidentiality of other data (e.g., secret keys). However, secret keys can be compromised. For instance, Intel SGX-based trusted execution environments have recently been compromised because the root provisioning key (SGX Fuse Key0) and the root sealing key (SGX Fuse Key1) can be leaked.\footnote{\url{https://x.com/PratyushRT/status/1828183761055330373}}

\example
We trust the compiler to be non-malicious. In his Turing award lecture \cite{trust}, Thompson presents a process to create an open-source compiler that can stealthily inject vulnerabilities into the compiled binary: In Stage 1, we write a compiler that can compile itself. In Stage 2, we add code to inject a vulnerability into the compile\textbf{d} binary and itself into the compile\textbf{r} binary. Then, we recompile the compiler. In Stage 3, we remove the offending code from the open-source compiler, recompile, and distribute the resulting compiler binary. The compiler binary will eternally inject the vulnerability-injecting code when compiling its own vulnerability-free source. This is Thompson's reflection on the trust we put into our tools and the people writing these tools.

\example
Project maintainers rely on a code review process to check the absence of bugs in new contributions. While we can safely assume that the vast majority of contributions are benign, without additional tooling there are opportunities for an attacker to exploit the developer's trust in code contributions and inject vulnerabilities that could later be exploited \cite{insecurity}.\footnote{XZ open source attack (Jia Tan): \url{https://research.swtch.com/xz-timeline}}

\example
We trust that our automated tooling, including compilers, linkers, debuggers, fuzzers, static analysers, verifiers, and theorem provers, work as advertised. For instance,
\begin{itemize}[leftmargin=0.3cm,itemsep=0.05cm]
\item Even if we trust the security policies of the linux kernel, there may be bugs in the \emph{eBPF verifier}\footnote{eBPF (\url{https://eBPF.io}) enables users to instrument a running system by loading small programs into the operating system kernel. It is used, e.g., to implement security policies.} preventing us from detecting policy violations \cite{ebpfVerif,ebpfVerif3,ebpfVerif2}.
\item Even if we trust the fuzzer, there may be bugs in the \emph{sanitizer}  which would prevent the fuzzer from flagging security flaws that are actually in scope \cite{{ubfuzz}}.
\item Even if we trust the software verifier, there may be bugs in the \emph{constraint solver} that is used which could undermine the provided formal guarantee \cite{fuzzsmt2, fuzzsmt1}.
\end{itemize}

\vspace{0.1cm}
\noindent
\emph{Problem of Infinite Regress} \cite{sep-infinite-regress}.
In epistemic philosophy, this is the problem of Infinite Regress: A belief is justified because it is based on another belief that is justified. We must anchor our trust somewhere. While these trust anchors are often well-specified in security, they are not exempt from scrutiny and subject to the same challenges we have listed in support of our theory. An attacker may be able to attack the system by exploiting bugs or security flaws in our trust anchors.

\subsubsection{Competing points of view.}\label{sec:view}
Security is a primary concern only for us.
There are many stakeholders in a software system. These stakeholders have different \emph{perspectives} on the system that might (often unintentionally) not include security. Stakeholders often have different \emph{requirements} for the software system which are sometimes conflicting with its security.
An attacker may exploit design decisions made to accommodate non-security preferences of other stakeholders.

\example
While the developer takes a \emph{constructive perspective}, the attacker takes an \emph{adversarial perspective} \cite{adversarialThinking}. On the one hand, the \emph{developer} reads code to implement \emph{intended} features and to fix bugs. On the other hand, the \emph{attacker} reads code (or binary) to exploit \emph{unintented} features (i.e., the weird machine) \cite{bratus,dullien,vanague}. These conflicting perspectives must be reconciled. The developer implements with intention but might realize something that deviates from this intention \cite{deviant}. This difference between intention and realization can be exploited ``to program the weird machine" \cite{dullien}.

\example
\emph{Better performance} often wins against better security. For instance, the browser security team of Microsoft Edge recently found that disabling the Just-In-Time (JIT) optimization could mitigate half of the Chrome exploits observed in-the-wild.\footnote{\url{https://microsoftedge.github.io/edgevr/posts/Super-Duper-Secure-Mode/}} In fact, performance optimization is a root cause of many critical types of security flaws. For instance, \emph{memory corruption} (such as buffer overflow or use-after-free) is often caused by undefined behavior which is deliberately left undefined in the C language specification for optimization reasons. \emph{Side channel attacks} are often enabled by microarchitectural or compiler optimizations. For some critical types of security flaws, \emph{we even have mitigations}. However, for performance reasons they are often left disabled in production.

\example
Competing points of view might result in stakeholder requirements that conflict with security.
\begin{itemize}[leftmargin=0.3cm,itemsep=0.1cm]
  \item The \emph{vendor} cares about time-to-market where the software system is tested in production and new features are deployed almost instantly.
  \item The \emph{customer} cares about a low cost. In fact, Woods~\cite{woods} suggests if buyers cannot distinguish secure from insecure product, then there is incentive to sell insecure products. Software security exists within a \emph{lemons market} \cite{lemons}.
  \item The \emph{user} cares about performance and a nice user experience, such that the performance overhead of certain security measures is considered impractical.
  \item The \emph{developer} cares about understanding the source code, such that certain side effects to make code more secure (e.g., more verifiable) are not appreciated \cite{secdev18, assertion2,denning99}.
\end{itemize}

\example
Some maintainers remain unconvinced that rewriting components of their project in a memory-safe langauge will help improve the security of the project:
\said{``Will I ever rewrite curl in rust? I don't believe in rewrites, no matter which language. I believe in replacing code and fixing components gradually over time. That *could* mean that we have a curl written mostly in rust in 10 years. Or in 20 years. Or not." 
\begin{flushright}---\emph{Daniel Stenberg (curl main developer) \cite{curl}}\end{flushright}
}

When developing and advocating our software security tooling and processes, we should always consider the different stakeholders and their (potentially competing) perspectives. However, in general competing requirements practically prevent us from providing guarantees about the software system.

\subsubsection{Security is a function of incentive.}\label{sec:incentive}
Security is a function of incentive. Without incentive to attack, a system may only appear to be secure, until there is incentive. The incentive is higher the more widely used or the more critical the software system, component, or library is. The incentive can also be artificially generated via red teaming, bug bounty programs, or pwning competitions.
Irrespective of the tools, techniques, and processes we have in place to maximize the security of our system---if there is no real (or artificial) incentive to develop exploits and to report these bugs, we will never know about the true insecurity of the system---irrespective of our confidence in its security---\emph{until} there is incentive.

\example
The more incentive there is for ethical hackers to find and report security flaws, the more independent scrutiny this system will have undergone.\footnote{This also means that the number of security flaws that are \emph{known} for this system might be a measure more of its criticality than its insecurity.}
Apple has become one of the most innovative companies in terms of mitigations (KIP, SCIP, PAC, PPL). For several years, despite the hard work and the awesome innovations of the security team at Apple, it took no more than a few weeks from the release of the new iPhone to the next jailbreak. The large market for jailbreaks has provided a strong incentive for hackers. However, when two iOS researchers presented the technical details of the most recent jailbreak at NullCon Goa 2022 \cite{jbtihm}, they responded to a question on how to get started learning to jailbreak: ``It is already too late'': While a kernel read/write had been sufficient a few years prior, it was only the starting point eight months before their successful jailbreak that year.

\emph{Mitigations as cost}. Any mechanism or process to improve the security of a system also increases the cost which subtracts from the incentive. Due to this incentive structure, an attacker will always target low hanging fruits first. Once a mitigation is adopted to prevent a  vulnerability class \emph{currently} popular, attackers will just move on to the next low hanging fruit \cite{blackhat16}.

\emph{Security-by-obscurity not viable}. According to Kerckhoff's principle, a system should be secure \emph{even \textbf{if}} everything about the system is public knowledge. Obfuscation and other deterrents increase the imbalance between attacker and defender in favor of the attacker. While there is obviously no incentive for ethical hackers to discover and responsibly disclose existing security flaws, a determined attacker can be expected to have more skills, resources, and incentive to exploit the (unhandled) security flaws despite the deterrent.
The security of a system should not depend on the secrecy of its components.
The reputational and financial damage arising from unknown exploits is significantly worse than inviting responsible disclosure via an open security approach.

\example
Dellago et al. \cite{exploitbrokers} studied the market for 0-day exploits from 2016 and 2021.\footnote{0-day exploits use vulnerabilities that are unknown to the developers.} For the four studied operating systems (Windows, MacOS, iOS, and Android), the number of 0-days observed in the wild decreased from 2016 to 2018 but increased again until the end of the study period. For the two investigated exploit brokers, Zerodium and Crowdfence, exploit prices increased to a maximum in 2019 and have since remained stable at USD 2.5-3 \emph{million} for a full exploit chain with persistence on Android or iOS. Recently, a third broker, Operation Zero, has announced to offer USD 20 \emph{million} for the same, which Google's Senior Director of the Threat Analysis Group, Shane Huntley, took as a good sign that the developed mitigations are becoming more effective and exploits more difficult.\footnote{\scriptsize\url{https://twitter.com/ShaneHuntley/status/1706944206521160070}} Schechter \cite{exploitbrokers2} confirms that higher prices are a signal of more secure products.

\emph{Economic and legal guarantees}.
Even if a software system is formally certified to be secure and no security flaw is \emph{known}, it might still ``become'' insecure once there is enough incentive to develop an attack. A lack of incentive or visibility practically prevents us from thoroughly assessing the security of a system.
An economic or legal framework can provide such incentives. For instance, a \emph{bug bounty program} guarantees (towards all stakeholders) that the software system is at least as secure as they would pay someone else for reporting a security flaw in the system \cite{exploitbrokers2,woods}. A bug bounty program also serves as a signal to the vendor about the security of the system and effectively improves security, too.


\section{A Philosophy of Security}\label{sec:roadmap}
While our \emph{theory of guarantees} is meant to \emph{explain} why we must generally assume that exploitable security flaws exist in any sufficiently complex and widely-used software system, our \emph{philosophy of security} is meant to \emph{serve as a guiding principle} in the acknowledgement of this conclusion.

\emph{Consequences}. What does our theory mean for cybersecurity as a research field? Firstly, our theory systematically elucidates the fundamental attack surface of \emph{any} software system which which, hence identified, can now be hardened in a concrete and directed manner.
Secondly, our theory explains why research in cybersecurity might seem reactive, specific, and ephemeral: Defenses are always built as a reaction to and specific to one type of attack, and vice versa.
Lastly, our theory motivates a new philosophy of security.

\emph{Our philosophy}. How to maxmize the reliability of our statements about the world in the absence of a reliable approach? In his philosophy of science, Popper suggests that a scientific theory must evolve in a counterexample-guided manner via \emph{falsification}~\cite{popper}. In his philosophy of maths, Lakatos suggests that even mathematical theorems evolve in a counterexample-guided manner via \emph{proofs and refutations}~\cite{lakatos}. Similarly, we suggest in our philosophy of security that claims of security must evolve in a counterexample-guided manner,~too.

\said{``For Lakatos, the development of mathematics should not be construed as series of deductions [..]. Rather, these water-tight deductions from well-defined premises are the (perhaps temporary) \textbf{end-points} of an evolutionary, and indeed a \textbf{dialectical}, process in which the constituent concepts are initially ill-defined, open-ended or ambiguous but become sharper and more precise in the context of a protracted debate.''
\begin{flushright}\footnotesize---\emph{Musgrave and Pigden (Stanford Encyclopedia of Philosophy) \cite{sep-lakatos}\\[-0.1cm]\tiny(emphasis ours)}\end{flushright}
} 

\emph{Problem of Induction} \cite{sep-induction-problem}.
The prevalent philosophy of security has been to develop security approaches with \emph{inductive} claims about their effectiveness, i.e., to demonstrate the \emph{generality} of the security approach: When a new security technique or process $T$ is developed, we deem it as effective if no security flaw is missed in an evaluation. However, such inductive, universal claims can never be confirmed despite all evidence in favor \cite{herley}. Just \emph{one} identifed security flaw becomes undisputable evidence against any claim of security that arises from $T$ \cite{bratus}.
In fact, our theory suggests that, fundamentally, \emph{no} such approach can exist. Hence, our philosophy suggests a different method with a focus is on identifying and remediating the \emph{specific limits} of a technique (which implements the security approach) instead of its generality. We can never claim that a technique is effective in absolute terms.

\subsection{Vulnerability-Guided Hardening of Security Approaches}
We propose a counterexample-guided improvement of our security techniques and processes to gradually maximize their effectiveness. We should always seek to find evidence \emph{against} the generality of the technique, and reconcile such counterexamples with that technique. Specifically, we suggest
\begin{enumerate}[itemsep=0.1cm,leftmargin=0.5cm]
  \item Implement a security approach $T$ to protect software $S$.
  \item Find evidence $ce$ \emph{against} the security of $S$ despite $T$ (e.g., from bug reports solicited via a bug bounty program).
  \item Patch $S$ and improve $T$ by (i)~debugging what made $T$ ineffective in the presence of $ce$, and (ii)~changing $T$ to make it effective for the most general version of $ce$.
  \item Go to (2).
\end{enumerate}
But how does this method address the identified challenges?

\emph{Operationalization} (\S\ref{sec:oper}). Our philosophy addresses the first challenge of finding which properties must hold in the first place (unknown unknowns) and of drawing the right line with the right pen by soliciting (and reacting to) in\-de\-pen\-dent vulnerability reports, e.g., via bug bounty programs, red teaming, auditing, or pwning competitions \cite{paul}. If a vulnerability exploits a behavior we have not previously considered as a violation of a security property or if it exploits an improper operationalization, we improve our techniques to detect, prevent, or mitigate similar types of vulnerabilities.
In this sense, our method offers an adversarial perspective \cite{adversarialThinking} on our security tooling and processes (in addition to the system itself).

\emph{Reasoning} (\S\ref{sec:reason}). Our philosophy addresses the challenge of reasoning about these operational properties at the right level of abstraction using empirically valid assumptions and a sufficiently expressive reasoning framework. Vulnerability discovery (Step~2) can be guided (i)~by systematically uncovering implicit and violating explicit assumptions and (ii)~by systematically exploring the gap between the software system and the abstractions used by our tools. If a discovered vulnerability exploits one of those those, (Step~3) we improve the technique by adjusting the assumptions and abstractions, or by improving the expressibility of the language, logic, or theory.
In this sense, our method offers systematic guidance for improving our security tooling and processes.

\emph{Sociotechnics} (\ref{sec:socio}). Our philosophy, and specifically soliciting independent vulnerability reports, addresses the remaining challenge of establishing trustworthiness and reconciling competing points of view. For instance, an effective bug bounty program exploits security as a function of incentive, does not require any assumption of trust, and signals to all stakeholders how much money can be offered for a vulnerability without getting any actual vulnerabilities reported. In fact, we could consider a bug bounty program as an \emph{economic guarantee of security} \cite{woods,exploitbrokers,exploitbrokers2}.
Once a vulnerability is reported, it can be used to harden our in-house security techniques and processes.

\subsection{Example: Counterexample-guided Meta Verification}\label{sec:metaverif}
To demonstrate the application of our vulnerability-guided hardening method, we instantiate it concretely for software verification which is traditionally presented as \emph{concluding with a proof of security}.
Instead of concluding with a proof, we propose to embed the proof process explicitly into a prove-and-break feedback loop. Successful attacks on the actual software system serve as counter-examples that must be systematically reconciled into the formalization. As this counterexample-guided approach allows us to verify the verification process itself, we call this approach as ``adversarial" verification or counterexample-guided \emph{meta verification}.\footnote{A related procedure is counterexample-guided abstraction refinement (CEGAR) \cite{cegar}. In meta verification, the counterexample violates the property at the highest level of \emph{refinement} (i.e., the deployed system) but not in the abstraction. In CEGAR, it is the other way around: the counterexample violates the property in the \emph{abstraction} but may not in a refinement.}

If a critical system can be attacked despite a formal proof of its security, what then is the utility of \mbox{provable security}? Formally verified software like SeL4 \cite{sel4} or Comp\-Cert \cite{compCert} are among the most secure, high-assurance systems to ever exist. While the process of proof already rids the program of many bugs \cite{pldi11}, the final proof inherently excludes large classes of security flaws and provides a precise formal guarantee. The provided guarantee holds within an explicit formalization and with respect to precisely stated assumptions and security requirements \cite{herley}.
Indeed, it is those artifacts whose empirical validity we propose to systematically scrutinize and harden.

\algrenewcommand\algorithmicrequire{\textbf{Input:}}
\algrenewcommand\algorithmicensure{\textbf{Output:}}
\begin{algorithm}[t]
\caption{Counterexample-guided Meta-Verification}
\label{alg:meta}
\begin{algorithmic}[1]\small
\Require Initial software system $S$
\Require Initial formalized assumptions $A$ and properties $P$
\Require Initial formal framework $F$
\State $\langle S, A, P, F\rangle = \text{\texttt{verify}}(S, A, P, F)$
\State $\text{Counterex. } ce = \text{\texttt{attack}}(S, A, P, F)$
\While{$ce$ \text{exists}}
  \State $\langle S, A, P, F\rangle = \text{\texttt{reconcile}}(ce, S, A, P, F)$
  \State $\langle S, A, P, F\rangle = \text{\texttt{verify}}(S, A, P, F)$
  \State $ce = \text{\texttt{attack}}(S, A, P, F)$
\EndWhile
\Ensure Verified software system $S$
\Ensure Updated assumptions $A$ and properties $P$
\Ensure Empirically more valid formal framework $F$
\end{algorithmic}
\end{algorithm}

Algorithm~\ref{alg:meta} shows our proposed counterexample-guided meta verification procedure. We invite the research community to develop systematic approaches for the individual functions. The procedure starts with the verification of the software system $S$ in Line~1, using an initial set of assumptions $A$ and properties $P$ formalized within the logic or theory of the verification framework~$F$.

However, in practice and in our proposed procedure, the verification (i.e., the function \texttt{verify}) does \emph{not end with the failure to verify}, but continues with systematic adjustments to $S$, $A$, $P$, or $F$ until the verification finally succeeds (returning an updated set $\langle S,A,P,F\rangle$). Once $S$ is successfully verified w.r.t. $A$ and $P$ within $F$, we suggest to \texttt{attack} $S$ to identify a security flaw $ce$ that still exists in the actual system $S$ \emph{despite} the successful verification of its formal model (Line~2). As long as we can find successful counterexamples $ce$, we must update the system and verification artifacts to \texttt{reconcile} the counterexample with the formal model of the system and re-verify using the updated verification artifacts (Lines~3--7).

It is interesting to note that meta verification effectively tackles the well-known argument of De Millo, Lipton, and Perlis \cite{demillo} against the utility of automated verification. The authors argue that the validity of a formal proof is usually derived within a \emph{social process}, and suggest that automated verification might remove this social process and thus the means by which to establish a proof's validity. Meta verification addresses their argument by substituting the social process with a programmatic falsification of the verification result until no more counterexample $ce$ can be found.

\subsubsection{Verification-guided Vulnerability Discovery.}
Once the software system has been fully verified, we suggest to try and attack the software system despite the formal proof of security. Apart from existing vulnerability discovery techniques (\S 3.2), we propose to systematically and perhaps automatically exploit the challenges of formal reasoning about software security so as to ``harden" the formal guarantee, the verification artifacts, and the software system.

To implement \texttt{attack}, we might identify black swan properties $p\not\in P$ or exploit an infidelity in the operationalization of some $p\in P$. For any assumption $a\in A$, we might seek to  invalidate $a$ for the actual system \cite{assumptions,assumptions2}. Given the formal framework $F$, we might seek to violate a property $p\in P$ for $S$ while keeping $p$ in the abstract model of $S$ in tact.
To increase public \emph{trust} in tooling and framework $F$, we propose to invite all stakeholders to subject $F$ itself to bug finding, and to maximize the layman's comprehension of the provided formal guarantees (or ``fine print" \cite{secdev18}).

\subsubsection{Counterexample-guided Reformalization.}
Once a violation $ce$ of a property $p\in P$ is identified in the deployed software system $S$ \emph{despite} the formal guarantees provided by the formal reasoning framework $F$ modulo the assumptions $A$, we must update the verification artifacts accordingly. To implement \texttt{reconcile}, we suggest to identify the verification artifacts $\langle A,P,F\rangle$ that are inconsistent with the counterexample and systematically and perhaps automatically update the artifact(s) to enforce a property violation in the presence of $ce$ for the unchanged $S$. For instance, we might add new properties to $P$, remove invalid assumptions from $A$, or update the reasoning framework $F$, e.g., by fixing a bug in the tooling \cite{fuzzsmt2} or by extending the reasoning rules (i.e., logic \cite{guarnieri} or assurance cases \cite{marsha}). Once the formalization recognizes $ce$, the software system $S$ should be fixed, re-verified, and attacked until no counterexample can be found (Lines~3--7; Alg.~\ref{alg:meta}).


\section*{Acknowledgments}

This paper was written out of a deep curiousity to understand the reliability of the statements that our approaches can generate about the properties of a software system, and how we should proceed if we accepted that these statements are fundamentally unreliable from a formal point of view. On the one hand, researchers perpetually promise guarantees and effectiveness. On the other hand, practitioners are troubled by the obvious lack of guarantees. This paper is an attempt to reconcile the promise with reality: \emph{\underbar{Even} a formal security proof must end with the invitation to find new vulnerabilities}.

We extend our special thanks to 
Toby Murray, Joshua J. Drake, Daniel Woods, Nathanial (d0nut) Lattimer, and Thomas (Halvar Flake) Dullien for the insightful feedback and inspiring discussions.

\bibliographystyle{IEEEtran}
\bibliography{references}

\end{document}